\documentclass{mem}
\usepackage{natbib}\usepackage{txfonts}\usepackage{balance}
\usepackage{graphicx}
\usepackage[a4paper,breaklinks,dvipdfm]{hyperref}
\idline{78}{1}
\begin{document}
\def\mpb{\smallskip\par\noindent$\bullet\,\,$}
\def\pn{\par\noindent}
\catcode`\@=11
\def\gsim{\ifmmode{\mathrel{\mathpalette\@versim>}}
    \else{$\mathrel{\mathpalette\@versim>}$}\fi}
\def\lsim{\ifmmode{\mathrel{\mathpalette\@versim<}}
    \else{$\mathrel{\mathpalette\@versim<}$}\fi}
\def\@versim#1#2{\lower 2.9truept \vbox{\baselineskip 0pt \lineskip
    0.5truept \ialign{$\m@th#1\hfil##\hfil$\crcr#2\crcr\sim\crcr}}}
\catcode`\@=12
\def\msun{\hbox{$M_\odot$}}
\def\lsun{\hbox{$L_\odot$}}

\title{
Rethinking Globular Clusters Formation
}

   \subtitle{}

\author
{Alvio Renzini\inst{}
}

\offprints{Alvio Renzini}

\institute{
Istituto Nazionale di Astrofisica --
Osservatorio Astronomico di Padova, 
Vicolo dell'Osservatorio 5,
I-35121 Padova,  Italy
\email{alvio.renzini@oapd.inaf.it}
}

\authorrunning{Alvio Renzini}

\titlerunning{Globular Cluster Formation}

\abstract{
This paper is aimed at emphasizing some of the main hints, constraints and difficulties we currently have in trying to understand how globular clusters formed, along with their multiple stellar
generations, an issue that must be regarded as intimately connected to the formation process itself.
Thus, the topics that are addressed include i) the required mass of the progenitor, ii) how to form 
new stars in an environment already crowded by a previous stellar generation, iii) how photometry
and spectroscopy appear to suggest different formation processes for second generation stars, iv) 
whether dilution with pristine material may (or may not) be necessary for the formation of second generations, v)
why the few clusters with multiple iron abundances are after all not so different from those that are homogeneous, and finally vi) why special environmental conditions may not be necessary for the formation of globular clusters with multiple stellar generations.

}
\maketitle{}

\section{Introduction}
After the discovery of the ubiquity and complexities of their
multiple generations (MG), we may just forget all that was said
before about Globular Cluster (GC) formation. Indeed, early classical attempts at making GCs 
\citep[e.g.,][]{pee68,fal85} where not yet exposed to the MG
issues, and were focused on a mass scale ($10^5-10^6\,\msun$) that now appears to dramatically underestimate the sheer size of the phenomenon of GC formation. So, we need to start from scratch in our attempt of laying down a credible scenario in the perspective  of developing a full-fledged  theory of GC formation.

While the MG phenomenon is widespread, all clusters appear to be somewhat different in this respect, i.e., there are 
no identical twins and each GC has its own individual version of the MG pattern. Thus GCs  differ from each other in the relative proportions of their MGs as revealed by the multiple photometric sequences in their color-magnitude diagram and by the different patterns exhibited by the correlations/anticorrelations among the abundances of the CNO and proton-capture elements. This dictates a scenario 
which should accommodate a great variance in the details, such as the number of MGs in each clusters and the specific chemical composition of each of them.

Together with these chemical peculiarities being almost absent among stars in the Galactic field, all this means that the MG phenomenon must be regarded as intimately related to the formation process of GCs themselves. This is to say that the environment internal to the forming GC must be instrumental for the production of stars with the chemical abundance patterns typical of SG stars.  This simple fact may be sufficient to discriminate among different scenarios for the formation of MGs.

In this brief account of the facts and problems we are facing on GC formation, I will limit myself to consider a few issues that set important  constraints, or that will add others once new observations will elucidate 
aspects which are still ambiguous. These issues include:

\medskip
\mpb
The mass budget problem and the nature of \par the progenitor

\mpb
Forming second generations on top of the \par first ones

\mpb
Photometric discreteness vs Spectroscopic \par  continuity

\mpb
Dilution: where the diluting material would  \par come from? is it really needed? 

\mpb
Clusters with multiple [Fe/H]s 

\mpb
Multiple populations are not confined to \par metal poor clusters

\section{The mass budget problem \& the nature of the progenitor}

Only $\sim 5\%$ of the mass of a first generation (FG) of stars  comes out as mass return with the
right composition (He, p-process elements) to make the second generation (SG). Therefore, the mass of a GC progenitor (the giant cloud out of which it formed) must have been;

\begin{equation}
M_{\rm progenitor}  \simeq  M_{\rm SG(today)} \times 20 \times \epsilon^{-2}, 
\end{equation}
\pn
where $M_{\rm SG(today)}$ is the mass of the SG stars in a today GC and $\epsilon$ is the star formation efficiency, i.e., the fraction of the gas actually turned into stars. It comes to the square power as it applies once to the formation of the FG  and then also to the formation of the SG. With a {\it canonical} $10\%$ star formation efficiency, the multiplying factor to 
$M_{\rm SG(today)}$ becomes $\sim 2000$, and for a typical $M_{\rm SG(today)}\simeq 10^5\msun$ this implies 
a progenitor mass of $\sim 2\times 10^8\msun$. (This would climb even  to $\sim 4\times 10^9\msun$ in the case of the most massive Galactic GC, $\omega$ Cen. ) 

Multiplying this estimate by $\sim 200$, the number of MW GCs, one comes to an estimate of $\sim  4\times 10^{10}\msun$ for the amount of
gas needed to form all the MW globulars,  with
$\sim 4\times  10^9\msun$  of dissolved FG stars. This latter mass in stars should have been lost into the MW halo by the progenitors of the present day GCs, a mass indeed similar to that of the current MW halo. These numbers are clearly quite frightening, suggesting that GC formation was a far more titanic affair than ever conceived before.

Of course, there is an alternative: for some reason the star formation efficiency $\epsilon$ might have been $>>10\%$, i.e., much larger than the estimated value for star formation in the giant molecular clouds in the 
present-day Galactic disk. If $\epsilon$ was of the order of unity than the required gas mass budget drops by a factor of $\sim 100$, which illustrates the range of uncertainty we have to face. Clearly, $\epsilon$ values near the two extremes of this range entail vastly different scenarios for the formation of GCs, their progenitors, and the whole MW halo. For the time being, we don't know what was the value of $\epsilon$
when the first and second generations formed in GCs, and we have to leave our fantasy free to explore this whole range of uncertainty.

\section{Forming second generations on top of the first ones}

The typical central density in massive GCs with multiple populations is around  $\sim 10^5\msun\, {\rm pc}^{-3}$, corresponding to a number density of atoms of  $n \simeq 10^7$ cm$^{-3}$.
First and second generation stars have similar central densities and often the SG prevails.
For a Chabrier/Kroupa IMF 150 stars (more massive than 0.1 $\msun$)  are formed every $\sim 100\;\msun$ of gas 
going into stars, hence one has  $\sim 10^5$ stars per cubic parsec.  Now, the formation of the SG needs to wait $\sim 10^8$ years for enough AGB ejected
materials to accumulate, reach  $n \sim 10^7$ cm$^{-3}$ and proceed to form the SG.
Thus, SG star formation needs to take place in an environment already inhabited by an extremely
dense stellar system. Is this a different mode of star formation, compared to the case of a molecular cloud
virtually devoided of pre-existing stars? Is  the gas smart enough to find free interstices
between FG stars to form new stars from scratch? or does the gas prefer to fall on top of pre-existing stars,
using them as nuclear seeds to grow them into SG stars?

We may face a similar problem near the Galactic Center, where episodes of star formation currently take place  in an even denser stellar environment. However, the situation there is further complicated by the presence of the supermassive black hole, which has then posed the even moredifficult  problem of how to form new stars in its vicinity. Yet, star formation in the Galactic Center environment may give us some hint on how SG stars may have formed in GCs.

For the time being, we may proceed with a simple, {\it back of the envelope} calculation. During the gas accumulation phase which lasts $\sim 10^8$ years, FG stars may accrete gas from the ISM. Assuming this happens according to the famous Bondi formula, the accretion rate would be:
\begin{equation}
{\dot M} ={4\pi\rho G^2M^2\over v^3},
\end{equation} 
which gives $\sim 10^{-8}\msun/$yr for $n=10^7$, $M=0.5\,\msun$ and $v =10$ km/s.
So (if the Bondi formula applies) during waiting/accumulation phase a sizable mass
(a good fraction of a solar mass) might be accreted, and it looks like that pre-existing (FG) stars may work as seeds for the growth of SG
stars. If so, SG stars would be FG stars having accreted an amount of
mass comparable to that they were born with. 

Is such a scenario tenable? One problem is  how to maintain the observed FG/SG dichotomy, because some stars should avoid accretion completely (the surviving FG) while others should be dominated by it (the present SG), which is unlikely to happen. This seems to me a major obstacle for such an accretion scenario, suggesting that for some reason the Bondi formula may not apply to these circumstances. However, to some extent accretion is
likely to take place, and may help explain some of the chemical peculiarities of GC stars, as first explored by Franca D'Antona \citep{dan83}.

\section{Photometric discreteness vs Spectroscopic continuity}

As amply documented in the literature as well at this meeting, MGs manifest themselves as separate sequences in various  color magnitude diagrams, with the width of each them being consistent with being due purely to photometric errors. Conversely, as also amply documented in the literature and at this meeting, chemical correlations/anticorrelations such as the [Na/Fe] vs [O/Fe] plots appear to form a continuous sequence. The result is that photometric evidence suggests that star formation proceeded in two or more bursts, interleaved by gas accumulation phases without much star formation going on, whereas the spectroscopic evidence seems to suggest that star formation and gas accumulation coming from the previous generation(s) where concomitant, continuous processes. Thus, did the formation of SGs 
proceed in a continuous fashion, or through a series of two or more separate star formation episodes?

The two scenarios are clearly very different, and we need to decide whether to believe to the photometric
discreteness or to the apparent continuity of the spectroscopic evidence. By good fortune this is an issue that observations may not take much to solve. Indeed, a discrete set of [Na/Fe] and [O/Fe] values can mimic a continuous distribution when blurred by abundance measurement errors. At least in one case, the GC  M4, 
it is quite evident that the distribution is indeed bimodal rather than continuous \citep{mar11}. Thus, refining spectroscopic abundance measurements, e.g., increasing integration times and S/N ratios, may soon solve this issue at least for a representative sample of GCs.

\section{Dilution: Where the diluting material would  come from? is it really needed?}

Dilution of AGB ejecta with pristine material is widely invoked both to account for the {\it banana} shape to the [Na/Fe] vs [O/Fe] anticorrelation and because current sodium and oxygen AGB yields seem otherwise 
at variance with the observed chemical patterns.

This has raised the issue of how pristine material could have been left aside, escaping any significant contamination by the huge number of supernovae that the FG should have produced. Where was such
pristine material stored? No convincing scenario has yet emerged about this, perhaps with the exception of that in which intermediate mass close binaries have done the job, as to some extent they must have contributed somewhat from their common-envelope phases \citep{van12}. Would  this be enough?

But, above all, do we really need dilution? As mentioned above, we cannot exclude that the apparent continuous Na-O anticorrelation is actually the result of measurement errors. On the other hand,  
AGB chemical yields depend on at least 5 or 6 adjustable parameters and only a tiny fraction of this enormous parameter space has been explored so far. One should have been very lucky to have found  the right combination at once. This calls for more AGB models to be calculated, exploring further this  parameter space, aiming to find whether combinations of the  parameters exist that produce the desired result, rather than limit oneself to a small set of combinations that at first sight may  look plausible.
Actually, we may reverse the problem: the composition of SG stars may help us in a better understanding of the evolution itself of AGB stars!

\section{Clusters with multiple [Fe/H]s ($\omega$~Cen, M22, Terzan 5, ...)}

Almost all GCs with multiple populations appear to be homogeneous in iron, with only a few ($\omega$  Cen, M22, and Terzan 5) exhibiting different [Fe/H] values in each sub-population. One may think that these clusters experienced a dramatically different formation history compared to those that are homogeneous in iron. Yet, I believe that their formation was not so different from that of the other clusters.

For example, in $\omega$  Cen the main SG contains only $\sim 20\,\msun$ of  more
iron than the FG, and it is sufficient that only  $\sim 0.2\%$ of  the ejecta of the core-collapse supernovae  from the FG were retained, while 99.8\% of such ejecta were
lost by the progenitor of $\omega$  Cen. Thus, one needs a very small fraction of the SN ejecta to be
retained, and it is not such a  surprise if in most cases nothing is
retained at all! The difference is just between losing $\sim 100\%$ and {\it only} $\sim 99.8\%$ of the iron
produced by supernovae.
So, there cannot be much difference between multiple-iron and single-iron
clusters.

\section{Multiple populations are not confined to metal poor clusters (NGC 6388 \& 6441)}

One plausible, widely entertained scenario for the formation of GCs appeals to nucleated dwarf galaxy
precursors which are then tidally stripped upon being captured by the MW, then leaving the surviving nucleus as its GC remnant. Dwarf galaxies are metal poor, and therefore this scenario may fit well in the case of those GCs which are metal poor and display the multiple population phenomenon, which are the overwhelming majority.

However, two metal rich GCs, namely  NGC 6388 and NGC 6441, exhibit well developed SGs.  Their metallicity is $\sim 1/3$ solar, they belong to the bulge, and qualify as belonging to the red/metal rich GC family that 
cannot have formed within the quiet environment of dwarfs, but has instead formed in a
turbulent, clumpy disk or within the bulge itself, when it was forming some 10 Gyr ago, i.e., ``at $z\sim 2$” .
Yet, they also have multiple, helium-rich SG stars as most recently illustrated  by deep WFC3 observations \citep{bel13}. The inference is that multiple populations in GCs can  form irrespective of the
environment inhabited by the progenitor!  They do so in a metal poor environment that may be represented by early dwarf galaxies,
but they also form in the much more violent environment  prevailing in young, massive galaxies 
at the epoch of their bulge formation. Therefore, these two clusters pose highly constraining demands 
to any attempt at understanding GC formation.

\section{Conclusions}

GC formation through a series of stellar generation is an extremely complex phenomenon  that still
escapes our understanding.
Yet, Nature finds very easy to do it, in a
short series of successive bursts,
recycling the exhaust of the previous
generation(s), and in our Galaxy it did so in
perhaps $\sim 200$ different ways.
This is a  big challenge for us, one that triggers us to venture towards trying to solve the puzzle.

In the above I have tacitly assumed that the materials forming SG stars came from intermediate mass stars 
during their AGB phase. Far from being proved, this looks to me just the least implausible scenario currently at our disposal. The other option, whereby  SG stars would form in the extruding disk of rapidly rotating stars, does not require the GC environment for the production of SG stars: each individual star
would give birth to its own progeny, such as in the  {\it gemmation} reproduction of some primitive organisms, no matter whether in a proto-cluster or in isolation. Hence, this scenario would predict that Na-rich, O-poor stars would be equally common in GCs and in the general field, which is not the case.

In summary, the challenges we are facing in trying to figure out how GCs formed are manifold.
They start with the sheer size of the progenitor gas cloud out of which the first generation was formed.
We then need to decide whether the second generations formed through one or more subsequent bursts, each individual burst being chemically homogeneous, or whether star formation proceeded in a quasi-continuous fashion, while the chemical composition of the ISM was changing being continuously replenished by stellar mass loss from the first generation. This second option appears less likely, given the 
sharpness of the photometric sequences. The next question concerns the mode of star formation itself.
Indeed, second generations need to form in a volume already densely occupied by the first generation
stars. These may, or may not act as condensation seeds for the second generation stars and may or may not accrete significant amount of gas along with its chemical signatures. Given this mess, it is somewhat surprising that so little
heavy elements from supernovae are retained by GCs, even in the case of the most massive of them,
i.e., $\omega$ Cen. The clear evidence is that supernova {\it feedback} is very efficient in clearing GC progenitors of their whole residual gas and newly synthetized metals. This poses a formidable difficulty, I believe, to the ``dilution scenario'', in which one appeals to the mixing of AGB ejecta with pristine gas in order to reproduce the abundance ratios observed in second generation stars. Finally, we must devise a scenario in which GCs with multiple populations can form in quiet environments, such as dwarf galaxies at large, as well as within galactic bulges at the peak of their star formation activity. All this together, makes the formation of globular clusters one of the most fascinating astrophysical puzzles  we now have on the table.

\begin{acknowledgements}

I wish to warmly thank the Headquarters of the Subaru Telescope at Hilo and their Director Prof. Nobuo Arimoto for their kind hospitality while this paper was written and set up. I also like to acknowledge Franca D'Antona and all the many members of the team led by Giampaolo Piotto, for the many fruitful discussions  we had on these matters, and for their contagious enthusiasm with which they pursue cutting-edge research on globular clusters. I also acknowledge support from the PRIN INAF 2009 ``Formation and Early Evolution of Massive Star Clusters”.

\end{acknowledgements}

\bibliographystyle{aa}

\end{document}